\documentclass[aps,prl,twocolumn,showpacs,showkeys]{revtex4}

\usepackage{amssymb}
\usepackage{graphicx}
\usepackage{amsmath}
\usepackage{color}
\usepackage{units}

\begin{document}

\title{Shubnikov-de Haas oscillations and electronic correlations in the layered organic metal  $\kappa$-(BETS)$_2$Mn[N(CN)$_2$]$_3$}

\author{M. V. Kartsovnik$^1\footnote{mark.kartsovnik@wmi.badw.de}$}
\author{V. N. Zverev$^{2,3}$}
\author{W. Biberacher$^1$}
\author{S. V. Simonov$^3$}
\author{I. Sheikin$^4$}
\author{N. D. Kushch$^5$}
\author{E. B. Yagubskii$^5$}

\affiliation{$^{1}$Walther-Meissner-Institut, Bayerische Akademie der
Wissenschaften, Walther-Meissner-Strasse 8, D-85748 Garching, Germany}
\affiliation{$^{2}$Institute of Solid State Physics, Russian Academy of Sciences,
Academician Ossipyan Str. 2, Chernogolovka, 142432 Russia}
\affiliation{$^{3}$Moscow Institute of Physics and Technology, Institutskii
9, Dolgoprudny, 141700, Russia}
\affiliation{$^{4}$Laboratoire National des Champs Magn\'{e}tiques Intenses, CNRS,
INSA, UJF, UPS, F-38042 Grenoble Cedex 9, France}
\affiliation{$^{5}$Institute of Problems of
Chemical Physics, Russian Academy of Sciences, Academician Semenov avenue 1,
Chernogolovka, 142432, Russia}

\begin{abstract}
We present magnetoresistance studies of the quasi-two-dimensional organic conductor $\kappa$-(BETS)$_2$Mn[N(CN)$_2$]$_3$, where BETS stands for bis\-(ethylene\-dithio)\-tetra\-selena\-fulvalene. Under a moderate pressure of 1.4\,kbar, required for stabilizing the metallic ground state, Shubnikov - de Haas oscillations, associated with a classical and a magnetic-breakdown cyclotron orbits on the cylindrical Fermi surface, have been found at fields above 10\,T. The effective cyclotron masses evaluated from the temperature dependence of the oscillation amplitudes reveal strong renormalization due to many-body interactions. The analysis of the relative strength of the
oscillations corresponding to the different orbits and its dependence on magnetic field suggests an enhanced role of electron-electron interactions on flat parts of the Fermi surface.
\end{abstract}

\maketitle

\section{Introduction}

Since the Lifshitz-Kosevich theory \cite{lifs55} provided a basis for universal quantitative description of magnetic quantum oscillations, these effects became one of most popular experimental means of studying the Fermi surface properties of metals \cite{crac73,shoe84}. Besides traditional metals, quantum oscillations of magnetoresitance (Shubnikov - de Haas, SdH effect) and magnetization (de Haas - van Alphen effect) have recently proved extremely useful in exploring more complex topical materials  such as cuprate \cite{seba15,helm15} and iron-based superconductors \cite{shis10,walm13,carr11}, topological insulators \cite{qu10,anal10,ando13}, heavy fermion compounds \cite{tail88,shis05,jiao15}, and organic charge-transfer salts \cite{wosn96,kart04,kart05}. Here we report on an experimental study of the high-field interlayer magnetoresistance of the layered conductor $\kappa$-(BETS)$_2$Mn[N(CN)$_2$]$_3$, demonstrating the power of the SdH effect in exploring Fermi surface properties of a quasi-two-dimensional correlated electronic system.

The present compound belongs to the family of bifunctional organic charge-transfer salts, in which conducting and magnetic properties are formed by different electronic subsystems spatially separated on a subnanometer level. The electrical conductivity is provided by delocalized $\pi$ electrons of fractionally charged BETS donors arranged in two-dimensional (2D) sheets, whereas magnetic properties are dominated by localized $d$-electron spins of Mn$^{2+}$ in the insulating anionic layers \cite{kush08}. In addition to the interesting, still not understood crosstalk between the two subsystems
\cite{vyas11,vyas12a,vyas12b}, the narrow, half-filled conduction band is a likely candidate for a Mott instability \cite{zver10}. The material undergoes a metal-insulator transition at $\approx 21$\,K \cite{kush08,zver10}. The insulating ground state is very sensitive to pressure: under a quasi-hydrostatic pressure of about 1\,kbar it is completely suppresses, giving way to a metallic and even a superconducting state with $T_c \approx 5$\,K \cite{zver10}. A thorough knowledge of the Fermi surface and other basic properties of the normal-state charge carriers is certainly mandatory for understanding the interplay between the various instabilities of the normal metallic state.
To this end we have carried out high-field magnetoresistance studies of $\kappa$-(BETS)$_2$Mn[N(CN)$_2$]$_3$ under pressure $p = 1.4$\,kbar. This pressure drives the system in the metallic part of the phase diagram, however, not far away from the insulating domain. We have found SdH oscillations with two fundamental frequencies, indicating a Fermi surface consistent with the predictions of the band structure calculations \cite{zver10}. Our analysis of the oscillation behaviour suggests strong electron correlations which are considerably dependent on the inplane wave vector.

\section{Experimental}

Single crystals of $\kappa$-(BETS)$_2$Mn[N(CN)$_2$]$_3$ were grown electrochemically, as described elsewhere \cite{kush08} and had a shape of small plates with characteristic dimensions of  $\simeq 0.5\times0.3\times0.02$\,mm$^3$. The largest surface of the plate was parallel to the highly conducting BETS molecular layers, which is defined as the crystallographic  $bc$-plane.

Resistive measurements were done with a standard four-probe a.c. technique using a low-frequency ($f \sim 20$\,Hz) lock-in amplifier. Two contacts were attached to each of two opposite sample surfaces with conducting graphite paste in order to measure the interlayer resistance. The magnetoresistance measurements were done in the temperature range $(0.35 - 1.4)$\,K in magnetic fields of up to 29\,T generated by the 24\,MW resistive magnet at the LNCMI-Grenoble. A quasi-hydrostatic pressure of 1.4\,kbar was applied using the Cu-Be clamp cell with silicon oil as a pressure medium and with a manganin coil for pressure control. The samples were aligned with the normal to conducting layers (crystallographic $a$-axis) being parallel to the magnetic field.

\section{Results and discussion}
\begin{figure}[tb]
	\centering
		\includegraphics[width=0.45\textwidth]{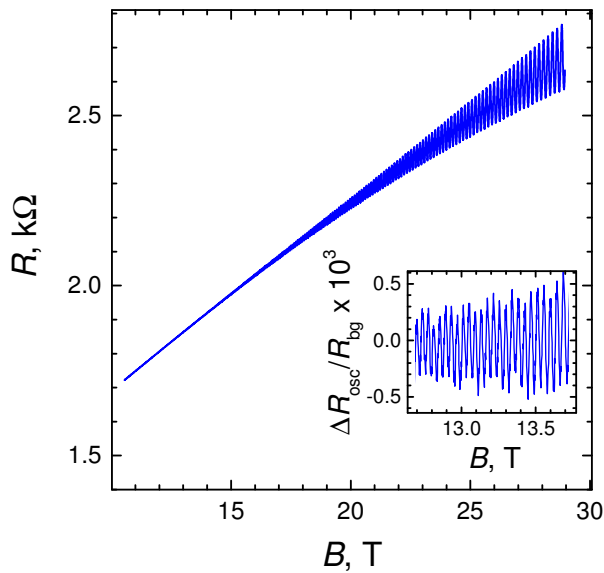}
	\caption{(Color online) Interlayer resistance of a $\kappa$-(BETS)$_2$Mn[N(CN)$_2$]$_3$ crystal as a function of magnetic field perpendicular to the layers, at $T= 0.4$\,K. Inset: close-up view of the oscillatory component of the field-dependent resistance normalized to the monotonic background: $\Delta R_{\mathrm{osc}}/R_{\mathrm{bg}} = R(B)/R_{\mathrm{bg}}(B)-1$. The background $R_{\mathrm{bg}}(B)$ was determined by a low-order polynomial fit to the as-measured resistance $R(B)$.}
	\label{SdH}
\end{figure}
\begin{figure}[tb]
	\centering
		\includegraphics[width=0.45\textwidth]{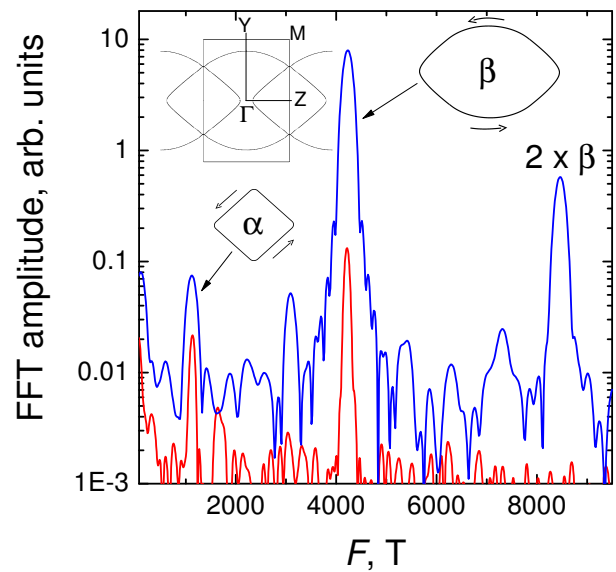}
	\caption{(Color online) FFT spectrum of the SdH oscillations at $T=0.4$\,K taken in the field windows $12 - 15$\,T (lower curve) and $23-29$\,T (upper curve). Two peaks at the fundamental frequencies $F_{\alpha}$ and $F_{\beta}$ correspond, respectively, to the classical ($\alpha$) and MB ($\beta$) orbits on the 2D Fermi surface \cite{zver10} shown in the upper left corner.}
	\label{FFT}
\end{figure}

Figure \ref{SdH} shows the low-temperature interlayer resistance of a pressurized $\kappa$-(BETS)$_2$\-Mn\-[N\-(CN)$_2$]$_3$ crystal measured in a magnetic field perpendicular to the layers. On the background of a monotonic, almost linear magnetoresistance one can see prominent SdH oscillations. The fast Fourier transform (FFT) of the oscillatory signal is presented in Fig.\,\ref{FFT} for two field windows,
12 to 15\,T and 23 to 29\,T. In both spectra the dominant frequency is $F_{\beta} = (4223 \pm 8)$\,T. In addition, a smaller peak is observed at $F_{\alpha} = (1126 \pm 8)$\,T and, in the higher-field spectrum, at $2F_{\beta} \approx 8460$\,T, which is the second harmonic of $F_{\beta}$. The contribution from the $\alpha$ frequency is stronger at relatively low fields, $< 20$\,T and can be seen by bare eye, for example, in the inset in Fig.\,\ref{SdH}. At increasing the field its relative contribution decreases; however, it is still present in the high-field FFT spectrum in Fig.\,\ref{FFT}.

The $\beta$ frequency reveals a cyclotron orbit area in $\mathbf{k}$-space equal to the first Brillouin zone (BZ) area. This result is in full agreement with the band structure calculations \cite{zver10}, predicting a 2D Fermi surface as shown in the upper left inset in Fig.\,\ref{FFT}. The corresponding cyclotron orbit in the extended zone scheme is shown in the upper right corner in Fig.\,\ref{FFT}.

Due to the inversion symmetry of the molecular layer, the band dispersion was predicted to be degenerate at the Fermi level on the line Z--M of the BZ boundary \cite{zver10}. In that case only one fundamental SdH frequency  $F_{\beta}$ would be expected. However, we clearly observe the frequency $F_{\alpha}$ corresponding to an orbit occupying $27\%$ of the BZ area. This is exactly the size of the $\alpha$ orbit centered at point Z of the BZ boundary, see Fig.\,\ref{FFT}. Therefore, we conclude that the band degeneracy is lifted, most likely due to a finite spin-orbit interaction \cite{wint_comm}. The resulting Fermi surface should consist of a closed part $\alpha$ and a pair of open sheets extended along $\mathbf{k_c}$ (parallel to $\Gamma$\,-- Z line in Fig.\,\ref{FFT}). The large orbit $\beta$ is then a consequence of magnetic breakdown (MB) through the small gap between the different parts of the Fermi surface. Similar evidence of the MB effect were found on several well known $\kappa$-(BEDT-TTF)$_2$X salts \cite{kart95c,balt96,harr98,weis99b}.

The SdH spectrum in Fig.\,\ref{FFT} differs from that reported earlier for this compound at similar pressures \cite{zver10}. In the earlier experiment no $F_{\alpha}$ and $F_{\beta}$ but, instead, very slow oscillations with the frequency $F_{\gamma} = 88$\,T were found. While the absence of the $\alpha$ and $\beta$ oscillations can easily be attributed to the higher temperatures and lower fields used in Ref.\,\citenum{zver10}, the absence of the slow oscillations in the present experiment is most likely caused by a different pressurizing procedure. In the previous work the samples were cooled at ambient pressure and then a pressure of $\sim 1$\,kbar was applied at temperatures below 20\,K, using the He-gas pressure technique. During the ambient pressure cooling, a superstructure transition was detected at $T = 102$\,K \cite{zver10}. This transition was proposed to give rise to small Fermi pockets responsible for the slow oscillations. By contrast, in the present work we apply pressure already at room temperature, using the clamp cell technique. The pressure probably prevents the formation of superstructure and the associated reconstruction of the Fermi surface. Indeed, the characteristic transition feature found in the ambient-pressure $R(T)$ cooling curves has not been detected under pressure.

We now turn to quantitative analysis of the oscillations. Generally speaking, the quasi-2D character of the electronic system may lead to strong violations of the standard Lifshitz-Kosevich (LK) theory, see, e.g., Refs.\,\citenum{cham02,grig03,gvoz04,kart04,kart05,grig12}. However, if we compare the present oscillations with those observed on some other highly 2D, clean organic metals \cite{lauk95,sand96,balt02,wosn01a}, their amplitude is relatively weak and the harmonic content is low. As will be shown below, the oscillations exhibit the conventional exponential temperature and field dependence. Therefore, in the following we apply the conventional LK formalism described in detail in Ref.\,\citenum{shoe84}. This approach has been proved to give reasonable results for other similarly anisotropic organic metals at not too high magnetic fields \cite{wosn96,sing00,kart04,toyo07}.

We consider the relative amplitudes of the oscillations in the form \cite{comment_1st_harm}:
\begin{equation}
A_j = A_{0,j}R_{T,j}R_{\mathrm{D},j}R_{\mathrm{MB},j},
\label{LK}
\end{equation}
where the subscript $j =\alpha,\beta$ labels the relevant orbit on the Fermi surface, $A_j = \Delta\sigma_{\mathrm{osc},j}/\sigma_{\mathrm{bg},j}$ is the amplitude of the oscillations in conductivity normalized to the respective nonoscillating background, $A_{j,0}$ is the field- and temperature-independent prefactor and $R_T$, $R_{\mathrm{D}}$, and $R_{\mathrm{MB}}$ are the damping factors caused, respectively, by finite temperature, scattering, and MB effects.

\begin{figure}[tb]
	\centering
		\includegraphics[width=0.45\textwidth]{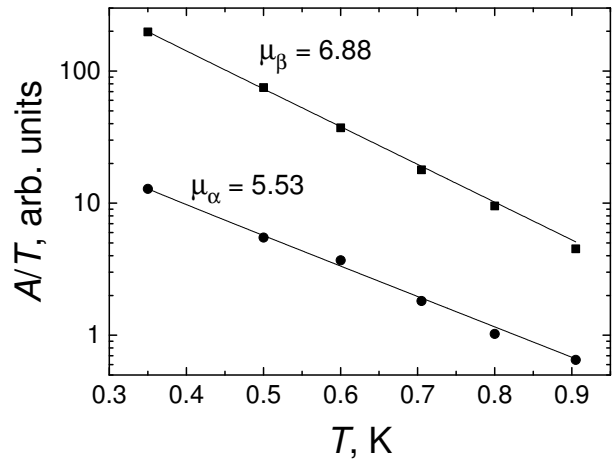}
	\caption{LK plot of the oscillation amplitudes $A_{\alpha}$ (circles) and $A_{\beta}$ (squares). The amplitudes were determined by FFT in the field window 14 to 17\,T. The lines are fits by Eq.\,(\ref{RT}) with normalized cyclotron mass values of 5.53 and 6.88 for the $\alpha$ and $\beta$ oscillations, respectively.}
	\label{mass}
\end{figure}

The temperature dependence of the SdH amplitude can be fitted by the LK temperature damping factor \cite{lifs55}:
\begin{equation}
R_T = \frac{K\mu T/B}{\sinh(K\mu T/B)}\,,
\label{RT}
\end{equation}
where $K = 14.69$\,T/K and $\mu = m_c/m_e$ is the effective cyclotron mass in units of the free electron mass $m_e$. For a large argument of the $\sinh$-function in Eq.\,(\ref{RT}) the logarithmic plot of the ratio $A_j/T(T)$, known as the LK plot, should be a straight line with a slope proportional to $\mu_j$. This is, indeed, true for both the $\alpha$ and $\beta$ oscillations, as demonstrated in Fig.\,\ref{mass}. Fitting the slopes yields the cyclotron masses $\mu_{\alpha} = 5.55 \pm 0.05$ and $\mu_{\beta} = 6.90 \pm 0.05$.

The obtained values can be compared to the band masses estimated within a noninteracting electron model. To this end, we have carried out tight-binding band structure calculations based on
the organic donor HOMOs (highest occupied molecular orbitals) obtained by the extended H\"{u}ckel method \cite{mori84a}. To estimate the energy dependence of the Fermi surface area and thus the cyclotron mass, the calculations were done for different band fillings near the Fermi level with a step of $0.001$ electron per BETS donor. The resulting values are $\mu_{\alpha 0} \simeq 1.1$ and $\mu_{\beta 0} \simeq 1.7$. Although these estimations are rather rough, they demonstrate that the real cyclotron masses are considerably enhanced, most likely due the many-body effects. The significance of electronic correlations would not be very surprising, since $\kappa$-(BETS)$_2$Mn[N(CN)$_2$]$_3$ is proposed to be a Mott insulator at ambient pressure \cite{zver10} and at the given pressure, $p = 1.4$\, kbar, the compound is still quite close to the metal-insulator boundary. It should be noted that the present mass values, especially $\mu_{\alpha}$, are even higher than those reported for the similar $\kappa$-(BEDT-TTF)$_2$X salts \cite{andr91,caul93,kart95c,weis99b,ohmi98}, which are also known for a strong Mott instability \cite{toyo07,kano04}.

While the temperature dependence of the oscillation amplitude is solely determined by the temperature damping factor $R_T$, the field dependence is contributed by all three damping factors on the right-hand side of Eq.\,(\ref{LK}). The Dingle factor taking into account the Landau level broadening $\Gamma$ has the form \cite{ding52,bych61a}:
\begin{equation}
R_{\mathrm{D}} = \exp\left( -2\pi \Gamma /\hbar \omega_{c}  \right) = \exp \left( -K\mu T_{\mathrm{D}}/B \right),
\label{RD}
\end{equation}
where $\omega_c = eB/m_c$ is the cyclotron frequency and $T_{\mathrm{D}} = \Gamma/\pi k_{\mathrm{B}}$ is the Dingle temperature. The Dingle temperature is often associated with the scattering rate $1/\tau$ \cite{shoe84}: $T_{\mathrm{D}} = 2\pi /\hbar k_{\mathrm{B}}\tau$. Finally, the MB factors for the $\alpha$ and $\beta$ oscillations are determined, respectively, by the probability amplitudes of Bragg reflection or tunneling at the MB junctions \cite{shoe84,fali66} and can be expressed as
\begin{equation}
R_{\mathrm{MB},\alpha} = \left[1 - \exp\left( -B_0/B \right) \right]
\label{RMBa}
\end{equation}
and
\begin{equation}
R_{\mathrm{MB},\beta} = \exp\left( -2B_0/B \right),
\label{RMBb}
\end{equation}
where $B_0$ is the characteristic MB field.

\begin{figure}[tb]
	\centering
		\includegraphics[width=0.45\textwidth]{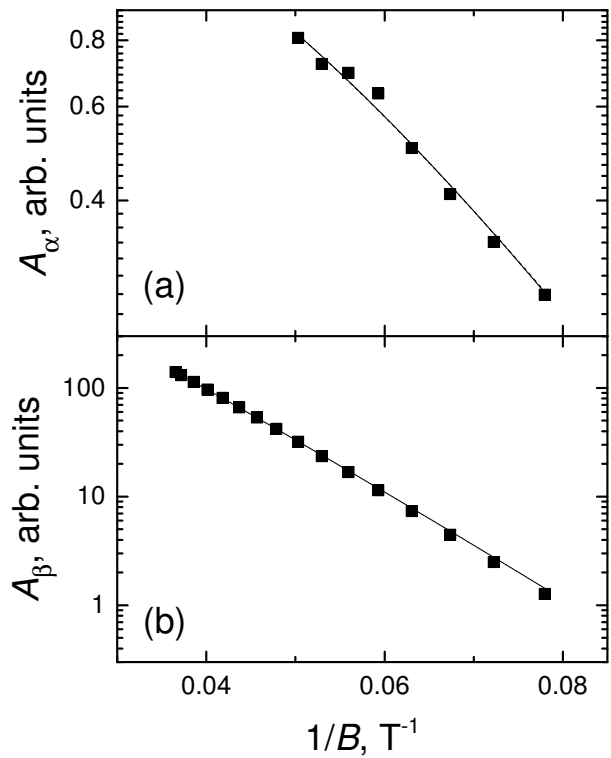}
	\caption{(a) Amplitude of the $\alpha$ oscillations as a function of inverse magnetic field; (b) the same for the $\beta$ oscillations. Both amplitudes are measured in the same units. The lines are fits by Eqs.\,(\ref{LK})-(\ref{RMBb}), see text.}
	\label{AB}
\end{figure}

Equations (\ref{LK}) - (\ref{RMBb}) can be used for analyzing the field dependence of the oscillation amplitudes. Fig.\,\ref{AB} shows the amplitudes of both oscillatory components plotted against inverse magnetic field. The data points are obtained from FFT spectra taken in 3\,T-wide field windows. The height of the $F_{\alpha}$ peak becomes comparable with the FFT background noise level above 20\,T; therefore, for this frequency only the data in the field range 12.8 to 20\,T is presented.

One readily sees from Eqs.\,(\ref{RD}) and (\ref{RMBb}) that $R_{\mathrm{MB},\beta}$ has the same functional dependence on $B$ as the Dingle factor. Therefore, the Dingle temperature and MB field cannot be separately determined from the field dependence of the $\beta$ oscillations, which dominate our SdH spectrum. On the other hand, the $B$-dependence of $R_{\mathrm{MB},\alpha}$ is rather weak as compared to the exponential behaviour of $R_{T,\alpha}$ and $R_{\mathrm{D},\alpha}$. This limits the accuracy of our analysis. Nevertheless, we will show that it enables us to draw some important qualitative conclusions.

For fitting we first assume that scattering is momentum-independent and, thus, the Dingle temperature is the same for both orbits. In this case, substituting the cyclotron masses determined above and performing an iterative fit of both $A_{\alpha}(B)$ and $A_{\beta}(B)$, we obtain $T_{\mathrm{D}} = 0.35$\,K and $B_0 \approx 26$\,T. Taking into account that, despite the heavier mass, the $\beta$ oscillations strongly dominate the SdH spectrum even at $B \sim 12-15$\,T, the obtained MB field seems to be much too high. Indeed, in order to match the observed relation between the amplitudes, we have to assume an unreasonably large ratio of the prefactors in Eq.\,(\ref{LK}): $A_{0,\beta}/A_{0,\alpha} =670$. Since the $\alpha$ orbit constitutes approximately one half of the $\beta$ orbit (see inset in Fig.\,\ref{FFT}), one would expect this ratio to be $\sim 2$. Of course, the ratio $A_{0,\beta}/A_{0,\alpha}$, being determined by the inplane momentum dependence of the Fermi surface parameters, needs not be exactly 2. Nevertheless, a modification by more than two orders of magnitude looks highly unlikely.

A key to resolving the apparent controversy is in a proper consideration of the influence of many body interactions. If the latter depend on the inplane momentum, the above assumption of a common Dingle temperature for both orbits is no longer valid. Although the rigorous analysis is very difficult, fortunately, the consideration may be strongly simplified by taking into account compensation of the renormalization effects on the cyclotron mass and Dingle temperature. For electron-phonon interactions such compensation was observed long ago on Hg \cite{pali72}: the Dingle factor $R_{\mathrm{D}}$ was found to be independent of temperature despite a considerable $T$-dependence of the electron-phonon scattering (see also Ref. \citenum{shoe84} for a review). It has been shown theoretically \cite{fowl65,enge70,elli80} that in a large field and temperature range the product $\mu T_{\mathrm{D}}$ can be approximated by the product of the bare cyclotron mass $\mu_0$ and Dingle temperature $T_{\mathrm{D}0}$ in the absence of electron-phonon scattering. While the case of electron-electron interactions is less studied in this respect, Martin et al. \cite{mart03} have proposed that the same compensation should hold for any inelastic processes, including the electron-electron scattering.

Turning to our compound, we note that the $\alpha$ pocket of the Fermi surface contains extended flat segments (see inset in Fig.\,\ref{FFT}) and hence has a pronounced nesting property. One can expect that electron correlations are particularly enhanced at the nesting wave vector. This is probably the reason for the unusually high effective cyclotron mass ratio $\mu_{\alpha}/\mu_{\beta} = 0.8$ as compared to the value $\approx 0.5$ typical of other $\kappa$-type organic conductors \cite{meri00a}. Indeed, the other salts do not have such flat segments and the interactions are believed to be momentum independent \cite{meri00a}. We, therefore, redo the analysis of the oscillations amplitudes in Fig.\,\ref{AB}, substituting the bare cyclotron masses $\mu_{\beta 0} =1.7$ and
$\mu_{\alpha 0} = \mu_{\beta 0}/2 = 0.85$ in the Dingle factor \cite{m_comm}. With an additional condition that the prefactor ratio $A_{0,\beta}/A_{0,\alpha} = 2$, the fit yields a reasonably small value of the MB field, $B_0 = 70$\,mT and only slightly different Dingle temperature values, 3.1\,K and 3.5\,K for the $\alpha$ and $\beta$ oscillations, respectively. Of course, the present analysis is far from being precise. An improvement can be achieved  when more reliable band mass estimations based on advanced, first-principle band structure calculations are available. Further theoretical and experimental studies are required for evaluation of the Fermi surface properties entering the prefactors $A_{0,j}$ in Eq.\,(\ref{LK}) for the oscillation amplitude.
Nevertheless, already now we can conclude that the behaviour of the SdH oscillations clearly reveals the importance of electron correlations in the present organic conductor.

\section{Acknowledgements}

We are grateful to S.M. Winter and P.D. Grigoriev for useful discussions.
The work was supported by the German Research Foundation (DFG) via the grant KA 1652/4-1.
The high-field measurements were done under support of the LNCMI-CNRS, member of the European
Magnetic Field Laboratory (EMFL). V.N.Z. acknowledges the support from RFBR Grant
15-02-02723.


\begin{thebibliography}{53}
\expandafter\ifx\csname natexlab\endcsname\relax\def\natexlab#1{#1}\fi
\expandafter\ifx\csname bibnamefont\endcsname\relax
  \def\bibnamefont#1{#1}\fi
\expandafter\ifx\csname bibfnamefont\endcsname\relax
  \def\bibfnamefont#1{#1}\fi
\expandafter\ifx\csname citenamefont\endcsname\relax
  \def\citenamefont#1{#1}\fi
\expandafter\ifx\csname url\endcsname\relax
  \def\url#1{\texttt{#1}}\fi
\expandafter\ifx\csname urlprefix\endcsname\relax\def\urlprefix{URL }\fi
\providecommand{\bibinfo}[2]{#2}
\providecommand{\eprint}[2][]{\url{#2}}

\bibitem[{\citenamefont{Lifshitz and Kosevich}(1956)}]{lifs55}
\bibinfo{author}{\bibfnamefont{I.~M.} \bibnamefont{Lifshitz}} \bibnamefont{and}
  \bibinfo{author}{\bibfnamefont{A.~M.} \bibnamefont{Kosevich}},
  \bibinfo{journal}{Sov. Phys. JETP} \textbf{\bibinfo{volume}{2}},
  \bibinfo{pages}{636} (\bibinfo{year}{1956}).

\bibitem[{\citenamefont{Cracknell and Wong}(1973)}]{crac73}
\bibinfo{author}{\bibfnamefont{A.~P.} \bibnamefont{Cracknell}}
  \bibnamefont{and} \bibinfo{author}{\bibfnamefont{K.~C.} \bibnamefont{Wong}},
  \emph{\bibinfo{title}{The Fermi Surface}} (\bibinfo{publisher}{Oxford
  University Press}, \bibinfo{address}{London}, \bibinfo{year}{1973}).

\bibitem[{\citenamefont{Shoenberg}(1984)}]{shoe84}
\bibinfo{author}{\bibfnamefont{D.}~\bibnamefont{Shoenberg}},
  \emph{\bibinfo{title}{Magnetic Oscillations in Metals}}
  (\bibinfo{publisher}{Cambridge University Press},
  \bibinfo{address}{Cambridge}, \bibinfo{year}{1984}).

\bibitem[{\citenamefont{Sebastian and Proust}(2015)}]{seba15}
\bibinfo{author}{\bibfnamefont{S.~E.} \bibnamefont{Sebastian}}
  \bibnamefont{and} \bibinfo{author}{\bibfnamefont{C.}~\bibnamefont{Proust}},
  \bibinfo{journal}{Annu. Rev. Condens. Matter Phys.}
  \textbf{\bibinfo{volume}{6}}, \bibinfo{pages}{411} (\bibinfo{year}{2015}).

\bibitem[{\citenamefont{Helm et~al.}(2015)\citenamefont{Helm, Kartsovnik,
  Proust, Vignolle, Putzke, Kampert, Sheikin, Choi, Brooks, Bittner
  et~al.}}]{helm15}
\bibinfo{author}{\bibfnamefont{T.}~\bibnamefont{Helm}},
  \bibinfo{author}{\bibfnamefont{M.~V.} \bibnamefont{Kartsovnik}},
  \bibinfo{author}{\bibfnamefont{C.}~\bibnamefont{Proust}},
  \bibinfo{author}{\bibfnamefont{B.}~\bibnamefont{Vignolle}},
  \bibinfo{author}{\bibfnamefont{C.}~\bibnamefont{Putzke}},
  \bibinfo{author}{\bibfnamefont{E.}~\bibnamefont{Kampert}},
  \bibinfo{author}{\bibfnamefont{I.}~\bibnamefont{Sheikin}},
  \bibinfo{author}{\bibfnamefont{E.-S.} \bibnamefont{Choi}},
  \bibinfo{author}{\bibfnamefont{J.~S.} \bibnamefont{Brooks}},
  \bibinfo{author}{\bibfnamefont{N.}~\bibnamefont{Bittner}},
  \bibnamefont{et~al.}, \bibinfo{journal}{Phys. Rev. B}
  \textbf{\bibinfo{volume}{92}}, \bibinfo{pages}{094501}
  (\bibinfo{year}{2015}).

\bibitem[{\citenamefont{Shishido et~al.}(2010)\citenamefont{Shishido, Bangura,
  Coldea, Tonegawa, Hashimoto, Kasahara, Rourke, Ikeda, Terashima, Settai
  et~al.}}]{shis10}
\bibinfo{author}{\bibfnamefont{H.}~\bibnamefont{Shishido}},
  \bibinfo{author}{\bibfnamefont{A.~F.} \bibnamefont{Bangura}},
  \bibinfo{author}{\bibfnamefont{A.~I.} \bibnamefont{Coldea}},
  \bibinfo{author}{\bibfnamefont{S.}~\bibnamefont{Tonegawa}},
  \bibinfo{author}{\bibfnamefont{K.}~\bibnamefont{Hashimoto}},
  \bibinfo{author}{\bibfnamefont{S.}~\bibnamefont{Kasahara}},
  \bibinfo{author}{\bibfnamefont{P.~M.~C.} \bibnamefont{Rourke}},
  \bibinfo{author}{\bibfnamefont{H.}~\bibnamefont{Ikeda}},
  \bibinfo{author}{\bibfnamefont{T.}~\bibnamefont{Terashima}},
  \bibinfo{author}{\bibfnamefont{R.}~\bibnamefont{Settai}},
  \bibnamefont{et~al.}, \bibinfo{journal}{Phys. Rev. Lett.}
  \textbf{\bibinfo{volume}{104}}, \bibinfo{pages}{057008}
  (\bibinfo{year}{2010}).

\bibitem[{\citenamefont{Walmsley et~al.}(2013)\citenamefont{Walmsley, Putzke,
  Malone, Guillamon, Vignolles, Proust, Badoux, Coldea, Watson, Kasahara
  et~al.}}]{walm13}
\bibinfo{author}{\bibfnamefont{P.}~\bibnamefont{Walmsley}},
  \bibinfo{author}{\bibfnamefont{C.}~\bibnamefont{Putzke}},
  \bibinfo{author}{\bibfnamefont{L.}~\bibnamefont{Malone}},
  \bibinfo{author}{\bibfnamefont{I.}~\bibnamefont{Guillamon}},
  \bibinfo{author}{\bibfnamefont{D.}~\bibnamefont{Vignolles}},
  \bibinfo{author}{\bibfnamefont{C.}~\bibnamefont{Proust}},
  \bibinfo{author}{\bibfnamefont{S.}~\bibnamefont{Badoux}},
  \bibinfo{author}{\bibfnamefont{A.}~\bibnamefont{Coldea}},
  \bibinfo{author}{\bibfnamefont{M.}~\bibnamefont{Watson}},
  \bibinfo{author}{\bibfnamefont{S.}~\bibnamefont{Kasahara}},
  \bibnamefont{et~al.}, \bibinfo{journal}{Phys. Rev. Lett.}
  \textbf{\bibinfo{volume}{110}}, \bibinfo{pages}{257002}
  (\bibinfo{year}{2013}).

\bibitem[{\citenamefont{Carrington}(2011)}]{carr11}
\bibinfo{author}{\bibfnamefont{A.}~\bibnamefont{Carrington}},
  \bibinfo{journal}{Reports on Progress in Physics}
  \textbf{\bibinfo{volume}{74}}, \bibinfo{pages}{124507}
  (\bibinfo{year}{2011}).

\bibitem[{\citenamefont{Qu et~al.}(2010)\citenamefont{Qu, Hor, Xiong, Cava, and
  Ong}}]{qu10}
\bibinfo{author}{\bibfnamefont{D.-X.} \bibnamefont{Qu}},
  \bibinfo{author}{\bibfnamefont{Y.~S.} \bibnamefont{Hor}},
  \bibinfo{author}{\bibfnamefont{J.}~\bibnamefont{Xiong}},
  \bibinfo{author}{\bibfnamefont{R.~J.} \bibnamefont{Cava}}, \bibnamefont{and}
  \bibinfo{author}{\bibfnamefont{N.~P.} \bibnamefont{Ong}},
  \bibinfo{journal}{Science} \textbf{\bibinfo{volume}{329}},
  \bibinfo{pages}{821} (\bibinfo{year}{2010}).

\bibitem[{\citenamefont{Analytis et~al.}(2010)\citenamefont{Analytis, McDonald,
  Riggs, Chu, Boebinger, and Fisher}}]{anal10}
\bibinfo{author}{\bibfnamefont{J.~G.} \bibnamefont{Analytis}},
  \bibinfo{author}{\bibfnamefont{R.~D.} \bibnamefont{McDonald}},
  \bibinfo{author}{\bibfnamefont{S.~C.} \bibnamefont{Riggs}},
  \bibinfo{author}{\bibfnamefont{J.-H.} \bibnamefont{Chu}},
  \bibinfo{author}{\bibfnamefont{G.~S.} \bibnamefont{Boebinger}},
  \bibnamefont{and} \bibinfo{author}{\bibfnamefont{I.~R.}
  \bibnamefont{Fisher}}, \bibinfo{journal}{Nat. Phys.}
  \textbf{\bibinfo{volume}{6}}, \bibinfo{pages}{960} (\bibinfo{year}{2010}).

\bibitem[{\citenamefont{Ando}(2013)}]{ando13}
\bibinfo{author}{\bibfnamefont{Y.}~\bibnamefont{Ando}},
  \bibinfo{journal}{Journal of the Physical Society of Japan}
  \textbf{\bibinfo{volume}{82}}, \bibinfo{pages}{102001}
  (\bibinfo{year}{2013}).

\bibitem[{\citenamefont{Taillefer and Lonzarich}(1988)}]{tail88}
\bibinfo{author}{\bibfnamefont{L.}~\bibnamefont{Taillefer}} \bibnamefont{and}
  \bibinfo{author}{\bibfnamefont{G.~G.} \bibnamefont{Lonzarich}},
  \bibinfo{journal}{Phys. Rev. Lett.} \textbf{\bibinfo{volume}{60}},
  \bibinfo{pages}{1570} (\bibinfo{year}{1988}).

\bibitem[{\citenamefont{Shishido et~al.}(2005)\citenamefont{Shishido, Settai,
  Harima, and \={O}nuki}}]{shis05}
\bibinfo{author}{\bibfnamefont{H.}~\bibnamefont{Shishido}},
  \bibinfo{author}{\bibfnamefont{R.}~\bibnamefont{Settai}},
  \bibinfo{author}{\bibfnamefont{H.}~\bibnamefont{Harima}}, \bibnamefont{and}
  \bibinfo{author}{\bibfnamefont{Y.}~\bibnamefont{\={O}nuki}},
  \bibinfo{journal}{J. Phys. Soc. Jpn.} \textbf{\bibinfo{volume}{74}},
  \bibinfo{pages}{1103} (\bibinfo{year}{2005}).

\bibitem[{\citenamefont{Jiao et~al.}(2015)\citenamefont{Jiao, Chen, Kohama,
  Graf, Bauer, Singleton, Zhu, Weng, Pang, Shang et~al.}}]{jiao15}
\bibinfo{author}{\bibfnamefont{L.}~\bibnamefont{Jiao}},
  \bibinfo{author}{\bibfnamefont{Y.}~\bibnamefont{Chen}},
  \bibinfo{author}{\bibfnamefont{Y.}~\bibnamefont{Kohama}},
  \bibinfo{author}{\bibfnamefont{D.}~\bibnamefont{Graf}},
  \bibinfo{author}{\bibfnamefont{E.~D.} \bibnamefont{Bauer}},
  \bibinfo{author}{\bibfnamefont{J.}~\bibnamefont{Singleton}},
  \bibinfo{author}{\bibfnamefont{J.-X.} \bibnamefont{Zhu}},
  \bibinfo{author}{\bibfnamefont{Z.}~\bibnamefont{Weng}},
  \bibinfo{author}{\bibfnamefont{G.}~\bibnamefont{Pang}},
  \bibinfo{author}{\bibfnamefont{T.}~\bibnamefont{Shang}},
  \bibnamefont{et~al.}, \bibinfo{journal}{Proc. Nat. Acad. Sci. USA}
  \textbf{\bibinfo{volume}{112}}, \bibinfo{pages}{673} (\bibinfo{year}{2015}).

\bibitem[{\citenamefont{Wosnitza}(1996)}]{wosn96}
\bibinfo{author}{\bibfnamefont{J.}~\bibnamefont{Wosnitza}}
  (\bibinfo{publisher}{Springer-Verlag}, \bibinfo{address}{Berlin Heidelberg},
  \bibinfo{year}{1996}).

\bibitem[{\citenamefont{Kartsovnik}(2004)}]{kart04}
\bibinfo{author}{\bibfnamefont{M.~V.} \bibnamefont{Kartsovnik}},
  \bibinfo{journal}{Chem. Rev.} \textbf{\bibinfo{volume}{104}},
  \bibinfo{pages}{5737} (\bibinfo{year}{2004}).

\bibitem[{kar()}]{kart05}
\bibinfo{note}{M. V. Kartsovnik and V. G. Peschansky, {\it Fiz. Nizk. Temp}
  {\bf 31}, 249 (2005) [{\it Low Temp. Phys.} {\bf 31}, 185 (2005)]}.

\bibitem[{\citenamefont{Kushch et~al.}(2008)\citenamefont{Kushch, Yagubskii,
  Kartsovnik, Buravov, Dubrovskii, Chekhlov, and Biberacher}}]{kush08}
\bibinfo{author}{\bibfnamefont{N.~D.} \bibnamefont{Kushch}},
  \bibinfo{author}{\bibfnamefont{E.~B.} \bibnamefont{Yagubskii}},
  \bibinfo{author}{\bibfnamefont{M.~V.} \bibnamefont{Kartsovnik}},
  \bibinfo{author}{\bibfnamefont{L.~I.} \bibnamefont{Buravov}},
  \bibinfo{author}{\bibfnamefont{A.~D.} \bibnamefont{Dubrovskii}},
  \bibinfo{author}{\bibfnamefont{A.~N.} \bibnamefont{Chekhlov}},
  \bibnamefont{and}
  \bibinfo{author}{\bibfnamefont{W.}~\bibnamefont{Biberacher}},
  \bibinfo{journal}{J. Am. Chem. Soc.} \textbf{\bibinfo{volume}{130}},
  \bibinfo{pages}{7238} (\bibinfo{year}{2008}).

\bibitem[{\citenamefont{Vyaselev et~al.}(2011)\citenamefont{Vyaselev,
  Kartsovnik, Biberacher, Zorina, Kushch, and Yagubskii}}]{vyas11}
\bibinfo{author}{\bibfnamefont{O.}~\bibnamefont{Vyaselev}},
  \bibinfo{author}{\bibfnamefont{M.}~\bibnamefont{Kartsovnik}},
  \bibinfo{author}{\bibfnamefont{W.}~\bibnamefont{Biberacher}},
  \bibinfo{author}{\bibfnamefont{L.}~\bibnamefont{Zorina}},
  \bibinfo{author}{\bibfnamefont{N.}~\bibnamefont{Kushch}}, \bibnamefont{and}
  \bibinfo{author}{\bibfnamefont{E.}~\bibnamefont{Yagubskii}},
  \bibinfo{journal}{Phys. Rev. B} \textbf{\bibinfo{volume}{83}},
  \bibinfo{pages}{094425} (\bibinfo{year}{2011}).

\bibitem[{\citenamefont{Vyaselev
  et~al.}(2012{\natexlab{a}})\citenamefont{Vyaselev, Kato, Yamamoto, Kobayashi,
  Zorina, Simonov, Kushch, and Yagubskii}}]{vyas12a}
\bibinfo{author}{\bibfnamefont{O.~M.} \bibnamefont{Vyaselev}},
  \bibinfo{author}{\bibfnamefont{R.}~\bibnamefont{Kato}},
  \bibinfo{author}{\bibfnamefont{H.~M.} \bibnamefont{Yamamoto}},
  \bibinfo{author}{\bibfnamefont{M.}~\bibnamefont{Kobayashi}},
  \bibinfo{author}{\bibfnamefont{L.~V.} \bibnamefont{Zorina}},
  \bibinfo{author}{\bibfnamefont{S.~V.} \bibnamefont{Simonov}},
  \bibinfo{author}{\bibfnamefont{N.~D.} \bibnamefont{Kushch}},
  \bibnamefont{and} \bibinfo{author}{\bibfnamefont{E.~B.}
  \bibnamefont{Yagubskii}}, \bibinfo{journal}{Crystals}
  \textbf{\bibinfo{volume}{2}}, \bibinfo{pages}{224}
  (\bibinfo{year}{2012}{\natexlab{a}}).

\bibitem[{\citenamefont{Vyaselev
  et~al.}(2012{\natexlab{b}})\citenamefont{Vyaselev, Kartsovnik, Kushch, and
  Yagubskii}}]{vyas12b}
\bibinfo{author}{\bibfnamefont{O.~M.} \bibnamefont{Vyaselev}},
  \bibinfo{author}{\bibfnamefont{M.~V.} \bibnamefont{Kartsovnik}},
  \bibinfo{author}{\bibfnamefont{N.~D.} \bibnamefont{Kushch}},
  \bibnamefont{and} \bibinfo{author}{\bibfnamefont{E.~B.}
  \bibnamefont{Yagubskii}}, \bibinfo{journal}{JETP Lett.}
  \textbf{\bibinfo{volume}{95}}, \bibinfo{pages}{565}
  (\bibinfo{year}{2012}{\natexlab{b}}).

\bibitem[{\citenamefont{Zverev et~al.}(2010)\citenamefont{Zverev, Kartsovnik,
  Biberacher, Khasanov, Shibaeva, Ouahab, Toupet, Kushch, Yagubskii, and
  Canadell}}]{zver10}
\bibinfo{author}{\bibfnamefont{V.~N.} \bibnamefont{Zverev}},
  \bibinfo{author}{\bibfnamefont{M.~V.} \bibnamefont{Kartsovnik}},
  \bibinfo{author}{\bibfnamefont{W.}~\bibnamefont{Biberacher}},
  \bibinfo{author}{\bibfnamefont{S.}~\bibnamefont{Khasanov}},
  \bibinfo{author}{\bibfnamefont{R.~P.} \bibnamefont{Shibaeva}},
  \bibinfo{author}{\bibfnamefont{L.}~\bibnamefont{Ouahab}},
  \bibinfo{author}{\bibfnamefont{L.}~\bibnamefont{Toupet}},
  \bibinfo{author}{\bibfnamefont{N.~D.} \bibnamefont{Kushch}},
  \bibinfo{author}{\bibfnamefont{E.~B.} \bibnamefont{Yagubskii}},
  \bibnamefont{and} \bibinfo{author}{\bibfnamefont{E.}~\bibnamefont{Canadell}},
  \bibinfo{journal}{Phys. Rev. B} \textbf{\bibinfo{volume}{82}},
  \bibinfo{pages}{155123} (\bibinfo{year}{2010}).

\bibitem[{win()}]{wint_comm}
\bibinfo{note}{S. Winter, private communication.}

\bibitem[{\citenamefont{Kartsovnik et~al.}(1995)\citenamefont{Kartsovnik,
  Biberacher, Andres, and Kushch}}]{kart95c}
\bibinfo{author}{\bibfnamefont{M.~V.} \bibnamefont{Kartsovnik}},
  \bibinfo{author}{\bibfnamefont{W.}~\bibnamefont{Biberacher}},
  \bibinfo{author}{\bibfnamefont{K.}~\bibnamefont{Andres}}, \bibnamefont{and}
  \bibinfo{author}{\bibfnamefont{N.~D.} \bibnamefont{Kushch}},
  \bibinfo{journal}{JETP Lett.} \textbf{\bibinfo{volume}{62}},
  \bibinfo{pages}{905} (\bibinfo{year}{1995}).

\bibitem[{\citenamefont{Balthes et~al.}(1996)\citenamefont{Balthes, Schweitzer,
  Heinen, Keller, Strunz, Biberacher, Jansen, and Steep}}]{balt96}
\bibinfo{author}{\bibfnamefont{E.}~\bibnamefont{Balthes}},
  \bibinfo{author}{\bibfnamefont{D.}~\bibnamefont{Schweitzer}},
  \bibinfo{author}{\bibfnamefont{I.}~\bibnamefont{Heinen}},
  \bibinfo{author}{\bibfnamefont{H.~J.} \bibnamefont{Keller}},
  \bibinfo{author}{\bibfnamefont{W.}~\bibnamefont{Strunz}},
  \bibinfo{author}{\bibfnamefont{W.}~\bibnamefont{Biberacher}},
  \bibinfo{author}{\bibfnamefont{A.~G.~M.} \bibnamefont{Jansen}},
  \bibnamefont{and} \bibinfo{author}{\bibfnamefont{E.}~\bibnamefont{Steep}},
  \bibinfo{journal}{Z. Phys. B} \textbf{\bibinfo{volume}{99}},
  \bibinfo{pages}{163} (\bibinfo{year}{1996}).

\bibitem[{\citenamefont{Harrison et~al.}(1998)\citenamefont{Harrison, Mielke,
  Rickel, Wosnitza, Qualls, Brooks, Balthes, Schweitzer, Heinen, and
  Strunz}}]{harr98}
\bibinfo{author}{\bibfnamefont{N.}~\bibnamefont{Harrison}},
  \bibinfo{author}{\bibfnamefont{C.~H.} \bibnamefont{Mielke}},
  \bibinfo{author}{\bibfnamefont{D.~G.} \bibnamefont{Rickel}},
  \bibinfo{author}{\bibfnamefont{J.}~\bibnamefont{Wosnitza}},
  \bibinfo{author}{\bibfnamefont{J.~S.} \bibnamefont{Qualls}},
  \bibinfo{author}{\bibfnamefont{J.~S.} \bibnamefont{Brooks}},
  \bibinfo{author}{\bibfnamefont{E.}~\bibnamefont{Balthes}},
  \bibinfo{author}{\bibfnamefont{D.}~\bibnamefont{Schweitzer}},
  \bibinfo{author}{\bibfnamefont{I.}~\bibnamefont{Heinen}}, \bibnamefont{and}
  \bibinfo{author}{\bibfnamefont{W.}~\bibnamefont{Strunz}},
  \bibinfo{journal}{Phys. Rev. B} \textbf{\bibinfo{volume}{58}},
  \bibinfo{pages}{10248} (\bibinfo{year}{1998}).

\bibitem[{\citenamefont{Weiss et~al.}(1999)\citenamefont{Weiss, Kartsovnik,
  Biberacher, Balthes, Jansen, and Kushch}}]{weis99b}
\bibinfo{author}{\bibfnamefont{H.}~\bibnamefont{Weiss}},
  \bibinfo{author}{\bibfnamefont{M.~V.} \bibnamefont{Kartsovnik}},
  \bibinfo{author}{\bibfnamefont{W.}~\bibnamefont{Biberacher}},
  \bibinfo{author}{\bibfnamefont{E.}~\bibnamefont{Balthes}},
  \bibinfo{author}{\bibfnamefont{A.~G.~M.} \bibnamefont{Jansen}},
  \bibnamefont{and} \bibinfo{author}{\bibfnamefont{N.~D.}
  \bibnamefont{Kushch}}, \bibinfo{journal}{Phys. Rev. B}
  \textbf{\bibinfo{volume}{60}}, \bibinfo{pages}{R16259}
  (\bibinfo{year}{1999}).

\bibitem[{\citenamefont{Champel}(2002)}]{cham02}
\bibinfo{author}{\bibfnamefont{T.}~\bibnamefont{Champel}},
  \bibinfo{journal}{Phys. Rev. B} \textbf{\bibinfo{volume}{65}},
  \bibinfo{pages}{153403} (\bibinfo{year}{2002}).

\bibitem[{\citenamefont{Grigoriev}(2003)}]{grig03}
\bibinfo{author}{\bibfnamefont{P.~D.} \bibnamefont{Grigoriev}},
  \bibinfo{journal}{Phys. Rev. B} \textbf{\bibinfo{volume}{67}},
  \bibinfo{pages}{144401} (\bibinfo{year}{2003}).

\bibitem[{\citenamefont{Gvozdikov}(2004)}]{gvoz04}
\bibinfo{author}{\bibfnamefont{V.~M.} \bibnamefont{Gvozdikov}},
  \bibinfo{journal}{Phys. Rev. B} \textbf{\bibinfo{volume}{70}},
  \bibinfo{pages}{085113} (\bibinfo{year}{2004}).

\bibitem[{\citenamefont{Grigoriev et~al.}(2012)\citenamefont{Grigoriev,
  Kartsovnik, and Biberacher}}]{grig12}
\bibinfo{author}{\bibfnamefont{P.~D.} \bibnamefont{Grigoriev}},
  \bibinfo{author}{\bibfnamefont{M.~V.} \bibnamefont{Kartsovnik}},
  \bibnamefont{and}
  \bibinfo{author}{\bibfnamefont{W.}~\bibnamefont{Biberacher}},
  \bibinfo{journal}{Phys. Rev. B} \textbf{\bibinfo{volume}{86}},
  \bibinfo{pages}{165125} (\bibinfo{year}{2012}).

\bibitem[{\citenamefont{Laukhin et~al.}(1995)\citenamefont{Laukhin, Audouard,
  Rakoto, Broto, Goze, Coffe, Brossard, Redoules, Kartsovnik, Kushch
  et~al.}}]{lauk95}
\bibinfo{author}{\bibfnamefont{V.~N.} \bibnamefont{Laukhin}},
  \bibinfo{author}{\bibfnamefont{A.}~\bibnamefont{Audouard}},
  \bibinfo{author}{\bibfnamefont{H.}~\bibnamefont{Rakoto}},
  \bibinfo{author}{\bibfnamefont{J.}~\bibnamefont{Broto}},
  \bibinfo{author}{\bibfnamefont{F.}~\bibnamefont{Goze}},
  \bibinfo{author}{\bibfnamefont{G.}~\bibnamefont{Coffe}},
  \bibinfo{author}{\bibfnamefont{L.}~\bibnamefont{Brossard}},
  \bibinfo{author}{\bibfnamefont{J.}~\bibnamefont{Redoules}},
  \bibinfo{author}{\bibfnamefont{M.}~\bibnamefont{Kartsovnik}},
  \bibinfo{author}{\bibfnamefont{N.}~\bibnamefont{Kushch}},
  \bibnamefont{et~al.}, \bibinfo{journal}{Physica B}
  \textbf{\bibinfo{volume}{211}}, \bibinfo{pages}{282} (\bibinfo{year}{1995}).

\bibitem[{\citenamefont{Sandhu et~al.}(1996)\citenamefont{Sandhu, Athas,
  Brooks, Haanappel, Goettee, Rickel, Tokumoto, Kinoshita, and
  Tanaka}}]{sand96}
\bibinfo{author}{\bibfnamefont{P.~S.} \bibnamefont{Sandhu}},
  \bibinfo{author}{\bibfnamefont{G.~J.} \bibnamefont{Athas}},
  \bibinfo{author}{\bibfnamefont{J.~S.} \bibnamefont{Brooks}},
  \bibinfo{author}{\bibfnamefont{E.~G.} \bibnamefont{Haanappel}},
  \bibinfo{author}{\bibfnamefont{J.~D.} \bibnamefont{Goettee}},
  \bibinfo{author}{\bibfnamefont{D.~W.} \bibnamefont{Rickel}},
  \bibinfo{author}{\bibfnamefont{M.}~\bibnamefont{Tokumoto}},
  \bibinfo{author}{\bibfnamefont{N.}~\bibnamefont{Kinoshita}},
  \bibnamefont{and} \bibinfo{author}{\bibfnamefont{Y.}~\bibnamefont{Tanaka}},
  \bibinfo{journal}{Surf. Sci.} \textbf{\bibinfo{volume}{361/362}},
  \bibinfo{pages}{913} (\bibinfo{year}{1996}).

\bibitem[{\citenamefont{Balthes et~al.}(2002)\citenamefont{Balthes, Wyder, and
  Schweitzer}}]{balt02}
\bibinfo{author}{\bibfnamefont{E.}~\bibnamefont{Balthes}},
  \bibinfo{author}{\bibfnamefont{P.}~\bibnamefont{Wyder}}, \bibnamefont{and}
  \bibinfo{author}{\bibfnamefont{D.}~\bibnamefont{Schweitzer}},
  \bibinfo{journal}{Solid State Commun.} \textbf{\bibinfo{volume}{124}},
  \bibinfo{pages}{141} (\bibinfo{year}{2002}).

\bibitem[{\citenamefont{Wosnitza et~al.}(2001)\citenamefont{Wosnitza, Wanka,
  Hagel, L\"{o}hneysen, Qualls, Brooks, Balthes, Schlueter, Geiser, Mohtasham
  et~al.}}]{wosn01a}
\bibinfo{author}{\bibfnamefont{J.}~\bibnamefont{Wosnitza}},
  \bibinfo{author}{\bibfnamefont{S.}~\bibnamefont{Wanka}},
  \bibinfo{author}{\bibfnamefont{J.}~\bibnamefont{Hagel}},
  \bibinfo{author}{\bibfnamefont{H.~V.} \bibnamefont{L\"{o}hneysen}},
  \bibinfo{author}{\bibfnamefont{J.~S.} \bibnamefont{Qualls}},
  \bibinfo{author}{\bibfnamefont{J.~S.} \bibnamefont{Brooks}},
  \bibinfo{author}{\bibfnamefont{E.}~\bibnamefont{Balthes}},
  \bibinfo{author}{\bibfnamefont{J.~A.} \bibnamefont{Schlueter}},
  \bibinfo{author}{\bibfnamefont{U.}~\bibnamefont{Geiser}},
  \bibinfo{author}{\bibfnamefont{J.}~\bibnamefont{Mohtasham}},
  \bibnamefont{et~al.}, \bibinfo{journal}{Phys. Rev. Lett.}
  \textbf{\bibinfo{volume}{86}}, \bibinfo{pages}{508} (\bibinfo{year}{2001}).

\bibitem[{\citenamefont{Singleton}(2000)}]{sing00}
\bibinfo{author}{\bibfnamefont{J.}~\bibnamefont{Singleton}},
  \bibinfo{journal}{Rep. Prog. Phys.} \textbf{\bibinfo{volume}{63}},
  \bibinfo{pages}{1111} (\bibinfo{year}{2000}).

\bibitem[{\citenamefont{Toyota et~al.}(2007)\citenamefont{Toyota, Lang, and
  M\"{u}ller}}]{toyo07}
\bibinfo{author}{\bibfnamefont{N.}~\bibnamefont{Toyota}},
  \bibinfo{author}{\bibfnamefont{M.}~\bibnamefont{Lang}}, \bibnamefont{and}
  \bibinfo{author}{\bibfnamefont{J.}~\bibnamefont{M\"{u}ller}}
  (\bibinfo{publisher}{Springer-Verlag Berlin Heidelberg},
  \bibinfo{year}{2007}), ISBN \bibinfo{isbn}{978 3 540 49574 1}.

\bibitem[{com()}]{comment_1st_harm}
\bibinfo{note}{Here we only consider the fundamental harmonics $F_{\alpha}$ and
  $F_{\beta}$}.

\bibitem[{\citenamefont{Mori et~al.}(1984)\citenamefont{Mori, Kobayashi,
  Sasaki, Kobayashi, Saito, and Inokuchi}}]{mori84a}
\bibinfo{author}{\bibfnamefont{T.}~\bibnamefont{Mori}},
  \bibinfo{author}{\bibfnamefont{A.}~\bibnamefont{Kobayashi}},
  \bibinfo{author}{\bibfnamefont{T.}~\bibnamefont{Sasaki}},
  \bibinfo{author}{\bibfnamefont{H.}~\bibnamefont{Kobayashi}},
  \bibinfo{author}{\bibfnamefont{G.}~\bibnamefont{Saito}}, \bibnamefont{and}
  \bibinfo{author}{\bibfnamefont{H.}~\bibnamefont{Inokuchi}},
  \bibinfo{journal}{Bull. Chem. Soc. Jpn.} \textbf{\bibinfo{volume}{57}},
  \bibinfo{pages}{627} (\bibinfo{year}{1984}).

\bibitem[{\citenamefont{Andres et~al.}(1991)\citenamefont{Andres, Heidmann,
  M\"uller, Himmelsbach, Biberacher, Probst, and Joss}}]{andr91}
\bibinfo{author}{\bibfnamefont{K.}~\bibnamefont{Andres}},
  \bibinfo{author}{\bibfnamefont{C.-P.} \bibnamefont{Heidmann}},
  \bibinfo{author}{\bibfnamefont{H.}~\bibnamefont{M\"uller}},
  \bibinfo{author}{\bibfnamefont{S.}~\bibnamefont{Himmelsbach}},
  \bibinfo{author}{\bibfnamefont{W.}~\bibnamefont{Biberacher}},
  \bibinfo{author}{\bibfnamefont{C.}~\bibnamefont{Probst}}, \bibnamefont{and}
  \bibinfo{author}{\bibfnamefont{W.}~\bibnamefont{Joss}},
  \bibinfo{journal}{Synth. Metals} \textbf{\bibinfo{volume}{41-43}},
  \bibinfo{pages}{1893} (\bibinfo{year}{1991}).

\bibitem[{\citenamefont{Caulfield et~al.}(1993)\citenamefont{Caulfield,
  Singleton, Pratt, Doporto, Lyubczynski, Hayes, Kurmoo, Day, Hendriks, and
  Perenboom}}]{caul93}
\bibinfo{author}{\bibfnamefont{J.}~\bibnamefont{Caulfield}},
  \bibinfo{author}{\bibfnamefont{J.}~\bibnamefont{Singleton}},
  \bibinfo{author}{\bibfnamefont{F.~L.} \bibnamefont{Pratt}},
  \bibinfo{author}{\bibfnamefont{M.}~\bibnamefont{Doporto}},
  \bibinfo{author}{\bibfnamefont{W.}~\bibnamefont{Lyubczynski}},
  \bibinfo{author}{\bibfnamefont{W.}~\bibnamefont{Hayes}},
  \bibinfo{author}{\bibfnamefont{M.}~\bibnamefont{Kurmoo}},
  \bibinfo{author}{\bibfnamefont{P.}~\bibnamefont{Day}},
  \bibinfo{author}{\bibfnamefont{P.~T.~J.} \bibnamefont{Hendriks}},
  \bibnamefont{and} \bibinfo{author}{\bibfnamefont{J.~A. A.~J.}
  \bibnamefont{Perenboom}}, \bibinfo{journal}{Synth. Met.}
  \textbf{\bibinfo{volume}{61}}, \bibinfo{pages}{63} (\bibinfo{year}{1993}).

\bibitem[{\citenamefont{Ohmichi et~al.}(1998)\citenamefont{Ohmichi, Ito,
  Ishiguro, Saito, and Komatsu}}]{ohmi98}
\bibinfo{author}{\bibfnamefont{E.}~\bibnamefont{Ohmichi}},
  \bibinfo{author}{\bibfnamefont{H.}~\bibnamefont{Ito}},
  \bibinfo{author}{\bibfnamefont{T.}~\bibnamefont{Ishiguro}},
  \bibinfo{author}{\bibfnamefont{G.}~\bibnamefont{Saito}}, \bibnamefont{and}
  \bibinfo{author}{\bibfnamefont{T.}~\bibnamefont{Komatsu}},
  \bibinfo{journal}{Phys. Rev. B} \textbf{\bibinfo{volume}{57}},
  \bibinfo{pages}{7481} (\bibinfo{year}{1998}).

\bibitem[{\citenamefont{Miyagawa et~al.}(2004)\citenamefont{Miyagawa, Kanoda,
  and Kawamoto}}]{kano04}
\bibinfo{author}{\bibfnamefont{K.}~\bibnamefont{Miyagawa}},
  \bibinfo{author}{\bibfnamefont{K.}~\bibnamefont{Kanoda}}, \bibnamefont{and}
  \bibinfo{author}{\bibfnamefont{A.}~\bibnamefont{Kawamoto}},
  \bibinfo{journal}{Chem. Rev.} \textbf{\bibinfo{volume}{104}},
  \bibinfo{pages}{5635} (\bibinfo{year}{2004}).

\bibitem[{\citenamefont{Dingle}(1952)}]{ding52}
\bibinfo{author}{\bibfnamefont{R.~B.} \bibnamefont{Dingle}},
  \bibinfo{journal}{Proc. Roy. Soc. A} \textbf{\bibinfo{volume}{211}},
  \bibinfo{pages}{517} (\bibinfo{year}{1952}).

\bibitem[{byc()}]{bych61a}
\bibinfo{note}{Yu. A. Bychkov, Zh. Eksp. Teor. Fiz. {\bf 39}, 1401-1410 (1960)
  [Sov. Phys. JETP {\bf 12}, 977-982 (1961)]}.

\bibitem[{\citenamefont{Falicov and Stachowiak}(1966)}]{fali66}
\bibinfo{author}{\bibfnamefont{L.~M.} \bibnamefont{Falicov}} \bibnamefont{and}
  \bibinfo{author}{\bibfnamefont{H.}~\bibnamefont{Stachowiak}},
  \bibinfo{journal}{Phys. Rev.} \textbf{\bibinfo{volume}{147}},
  \bibinfo{pages}{505} (\bibinfo{year}{1966}).

\bibitem[{\citenamefont{Palin}(1972)}]{pali72}
\bibinfo{author}{\bibfnamefont{C.~J.} \bibnamefont{Palin}},
  \bibinfo{journal}{Proc. R. Soc. A} \textbf{\bibinfo{volume}{329}},
  \bibinfo{pages}{17} (\bibinfo{year}{1972}).

\bibitem[{\citenamefont{Fowler and Prange}(1965)}]{fowl65}
\bibinfo{author}{\bibfnamefont{M.}~\bibnamefont{Fowler}} \bibnamefont{and}
  \bibinfo{author}{\bibfnamefont{R.~E.} \bibnamefont{Prange}},
  \bibinfo{journal}{Physica} \textbf{\bibinfo{volume}{1}}, \bibinfo{pages}{315}
  (\bibinfo{year}{1965}).

\bibitem[{\citenamefont{Engelsberg and Simpson}(1970)}]{enge70}
\bibinfo{author}{\bibfnamefont{S.}~\bibnamefont{Engelsberg}} \bibnamefont{and}
  \bibinfo{author}{\bibfnamefont{G.}~\bibnamefont{Simpson}},
  \bibinfo{journal}{Phys. Rev. B} \textbf{\bibinfo{volume}{2}},
  \bibinfo{pages}{1657} (\bibinfo{year}{1970}).

\bibitem[{\citenamefont{Elliott et~al.}(1980)\citenamefont{Elliott, Ellis, and
  Springford}}]{elli80}
\bibinfo{author}{\bibfnamefont{M.}~\bibnamefont{Elliott}},
  \bibinfo{author}{\bibfnamefont{T.}~\bibnamefont{Ellis}}, \bibnamefont{and}
  \bibinfo{author}{\bibfnamefont{M.}~\bibnamefont{Springford}},
  \bibinfo{journal}{J. Phys. F: Metal Phys.} \textbf{\bibinfo{volume}{10}},
  \bibinfo{pages}{2681} (\bibinfo{year}{1980}).

\bibitem[{\citenamefont{Martin et~al.}(2003)\citenamefont{Martin, Maslov, and
  Reizer}}]{mart03}
\bibinfo{author}{\bibfnamefont{G.~W.} \bibnamefont{Martin}},
  \bibinfo{author}{\bibfnamefont{D.~L.} \bibnamefont{Maslov}},
  \bibnamefont{and} \bibinfo{author}{\bibfnamefont{M.~Y.}
  \bibnamefont{Reizer}}, \bibinfo{journal}{Phys. Rev. B}
  \textbf{\bibinfo{volume}{68}}, \bibinfo{pages}{241309}
  (\bibinfo{year}{2003}).

\bibitem[{\citenamefont{Merino and McKenzie}(2000)}]{meri00a}
\bibinfo{author}{\bibfnamefont{J.}~\bibnamefont{Merino}} \bibnamefont{and}
  \bibinfo{author}{\bibfnamefont{R.~H.} \bibnamefont{McKenzie}},
  \bibinfo{journal}{Phys. Rev. B} \textbf{\bibinfo{volume}{62}},
  \bibinfo{pages}{2416} (\bibinfo{year}{2000}).

\bibitem[{m_c()}]{m_comm}
\bibinfo{note}{The substituted value of $\mu_{\alpha 0}$ is somewhat lower than
  obtained from our simplified band structure calculations; however, the mass
  ratio is consistent with the circumference ratio of the $\alpha$ and $\beta$
  orbits.}

\end{thebibliography}

\end{document}